\documentclass{emulateapj}
\usepackage{apjfonts}
\usepackage{natbib}
\usepackage{graphicx}
\usepackage{amssymb,amsmath}
\usepackage[export]{adjustbox}
\usepackage{xcolor}
\definecolor{ao}{rgb}{1.0,0.03,0}
\definecolor{ap}{rgb}{0, 0.18,0.39}
\usepackage[breaklinks,colorlinks,linkcolor=ao,citecolor=ap]{hyperref}

\shorttitle{CMB posterior using global ILC and Gibbs sampling}

\begin{document}

\title{An Application of Global ILC Algorithm over Large Angular Scales to Estimate  CMB Posterior 
Using Gibbs Sampling} 

\author{Vipin Sudevan\altaffilmark{1}, Rajib Saha\altaffilmark{1}}

\altaffiltext{1}{Physics Department, Indian Institute of Science 
Education and Research Bhopal,  Bhopal, M.P, 462023, India.}

\begin{abstract}
In this work, we formalize a new technique to investigate  joint posterior density 
of Cosmic Microwave Background (CMB) signal and its theoretical angular power spectrum
given the observed data, using the global internal-linear-combination (ILC) 
method first proposed by~\cite{Saha2017}. We implement the method on low resolution 
CMB maps observed by WMAP and Planck satellite missions,  using Gibbs sampling, assuming that the detector noise 
is negligible on large angular scales of the sky. The main products of our analysis 
are  best fit CMB cleaned map and its theoretical angular power spectrum along with 
their error estimates.  We validate 
the methodology by performing Monte Carlo simulations
that includes realistic foreground models and noise levels consistent with WMAP 
and Planck observations. Our method has an unique advantage that the posterior density 
is obtained without any need to explicitly model foreground components. Secondly, the power spectrum 
results with the error estimates can be directly used for cosmological parameter estimations. 

\end{abstract}


\section{Introduction}

Since the discovery of Cosmic Microwave Background (CMB)~\citep{P&W1965} rapid advancements in the field of its observation  
made it possible  to map the primordial signal over the entire sky with increasingly higher 
resolution~\citep{COBE1991, Bennett2013, Planck2018}. 
Accurate measurement of temperature (and polarization) anisotropy of CMB,  which arguably forms one of the 
cornerstones of the precision era of modern precision cosmology, provides us with a wealth of knowledge regarding the geometry, 
composition and the origin of the Universe (e.g., see~\cite{Planckcosmo2018} and references therein). However, 
the observed CMB signal in the microwave region is strongly contaminated due to
foreground emissions due to different astrophysical sources present within and 
outside our galaxy. Hence the challenge is to accurately recover the CMB signal for cosmological analysis,  by 
minimizing contributions from various foregrounds emissions.  

For reliable estimation of cosmological parameters, a desirable property of a CMB reconstruction method 
is that it produces both, the best guesses for the signal and its angular power spectrum
along with their corresponding error (and bias, if any) estimates.~\cite{Eriksen2004a, Eriksen2007, 
Eriksen2008, Eriksen2008a,PlanckCMB2016,PlanckFg2016} propose and implement a Gibbs sampling~\citep{Gibbs1984}
approach to jointly estimate the CMB map, its angular power spectrum and all foreground components
along with their error estimates using WMAP~\citep{Hinshaw2013}  and Planck~\citep{Planck2018} observations.    
\cite{Eriksen2006,Gold2011} use a maximum likelihood approach to reconstruct simultaneously CMB and foreground components 
using prior information about CMB and detector noise covariance matrices and foreground 
models. Although, these methods are extremely useful for simultaneous reconstruction of CMB and all 
foreground components, an alternative approach 
for CMB reconstruction alone, is the so-called internal-linear-combination 
(ILC) method~\citep{Tegmark96, Tegmark2003, Bennett2003,Eriksen2004, Saha2006} which does not
rely upon any explicit model of foreground spectrum. In recent years the method has been investigated 
extensively.~\cite{Saha2006} use this method to estimate 
CMB cross-power spectra by removing detector noise bias using WMAP maps. These authors also report 
presence of a possible negative bias at the low multipoles.~\cite{Saha2008}, for the first time, 
perform a rigorous analytical study  of negative  bias at the low multipoles for a single 
iteration ILC foreground removal procedure in harmonic space. Later~\cite{Sudevan2017} find and correct 
a foreground leakage in iterative ILC algorithm in harmonic space by applying their technique 
on high resolution Planck and WMAP observations.~\cite{Saha2017} propose a global ILC 
method in pixel space by taking into account prior information of CMB covariance matrix
under the assumption that detector noise can be ignored  over the large angular 
scales of the sky. The method considerably improves the usual ILC method at low resolution, 
where no prior information about the CMB covariance is used.  In spite of these progresses, a joint
analysis of CMB signal and its angular power spectrum posterior density in a foreground model
independent manner has not yet been explored in the literatures. The current article is aimed 
to provide a mechanism exactly to solve this problem. By estimating the posterior density of CMB signal
and CMB theoretical angular power spectrum given the observed  data over the large angular scales 
of the sky using the ILC method similar to~\cite{Saha2017},  we provide the 
best fit estimates of both, CMB map and theoretical angular power spectrum along with 
their  confidence interval regions. In the current article, we replace the CMB signal reconstruction technique  
 by a  faster harmonic domain algorithm than the pixel-space
algorithm of~\cite{Saha2017}.  We use 
Gibbs sampling method~\citep{GR1992} to draw samples from the joint conditional density. 
There are two important advantages of our method. First, the theoretical power spectrum results 
can directly be integrated to cosmological parameter estimation process. Second, 
the CMB posterior estimation can be achieved without any need to explicitly model the 
foreground components. The results, therefore, can not be sensitive to, foreground modeling 
uncertainties.  

The early work of CMB component reconstruction is performed  by~\cite{Bennett1992} using a variant 
of  ILC algorithm where prior information of free-free spectral index is used.~\cite{Bunn1994, Bouchet1999}
developed a Weiner filter approach.~\cite{Delabrouille2012, Delabrouille2013} propose an ILC
algorithm in needlet space, which can take into account local variation of foreground 
spectral properties both in the pixel and needlet space.~\cite{Saha2016} use the ILC method to jointly
reconstruct CMB Stokes Q polarization signal and other foreground components in presence of  
spatially varying spectral properties of polarized synchrotron emission using simulated observations 
of WMAP. In an interesting application of ILC method~\cite{Saha2011} and~\cite{Ujjal2017}              
reconstruct CMB maps using Gaussian nature of CMB and non-Gaussian nature of astrophysical foregrounds.

In Section \ref{formalism} we discuss the basic formalism. We describe the posterior estimation 
method in Section~\ref{Method}.  In Section \ref{Results} 
we present the results of analysis of WMAP and Planck frequency maps at low resolution. 
We discuss convergence tests of the Gibbs chains in Section~\ref{convergence}. 
We validate the posterior density estimation method by performing detailed Monte Carlo simulations 
using realistic foreground and detector noise model consistent with WMAP and Planck
observations in Section~\ref{MC}. Finally, we conclude in Section~\ref{Discussion}.

\section{Formalism}
\label{formalism}
\subsection{Data Model}

Let us assume that, we have  observations of foreground contaminated CMB maps at $n$  
different frequencies. Without sacrificing any generality, we assume that, each of these maps has the same beam 
(and pixel)  resolutions~\footnote{In general, different frequency maps have different beam resolutions. One 
can always bring these maps to a common beam resolution, as allowed by the experiment, by smoothing
by an appropriate kernel (e.g.,~\cite{Sudevan2017}). Same applies for the pixel resolution.}. The 
observed data set, ${\bf D}$, can be represented as ${\bf D} = \{{\bf X}_1, {\bf X}_2, ..., {\bf X}_n\}$, 
where $\textbf{X}_{i}$, $i \in \{1,2,...,n\}$, is an $N\times 1$ column vector denoting the input foreground 
contaminated CMB map (in thermodynamic temperature unit) at a frequency $\nu_i$. $N$ represents the number of 
pixels in each input frequency map and ${\bf D}$ is an $N \times n$ matrix. Assuming detector noise is negligible~\footnote{
We can safely assume this for WMAP and Planck temperature observations  on the large angular scales 
of the sky.} 
we have 
\begin{eqnarray}
{\bf X}_i  = {\bf S} + {\bf F}_i\, ,
\label{cmb_fg}
\end{eqnarray} 
where ${\bf S}$  is an $N\times 1$ column vector, representing the CMB signal~\footnote{CMB signal 
at any given direction on the sky is independent on frequency in thermodynamic temperature unit,
 since the former follows a blackbody spectrum to a very good accuracy.} and ${\bf F}_i$ denotes a map 
of same size representing  net foreground contamination at the frequency $\nu_i$. 

\subsection{CMB Posterior and Gibbs Sampling}
The CMB posterior 
density is denoted as $P({\bf S}, C_{\ell}| {\bf D})$, the joint density of CMB map, ${\bf S}$, and theoretical CMB angular power spectrum, 
$C_{\ell}$, given the observed data, ${\bf D}$.~\footnote{Since all  ${\bf X}_i$ and hence ${\bf S}$ inevitably 
contain some beam and pixel smoothing effects, we assume that,  the theoretical angular power spectrum $C_{\ell}$
also contain the same smoothing effects.}  A convenient way to establish the posterior density, without any need to
evaluate it,  is 
by drawing samples from the distribution itself. An useful sampling method in this  context is the so called 
Gibbs sampling approach~\citep{GR1992}, which states that the  posterior joint density under consideration conditioned on 
data can be established by following few steps. 
\begin{enumerate}

\item {Draw a sample, ${\bf S}^{i+1}$, from  from conditional density of CMB signal ${\bf S}$ 
given both, the  data ${\bf D}$ and some chosen  CMB theoretical angular power spectrum, $C^i_{\ell}$. 
Symbolically, 
\begin{equation}
{\bf S}^{i+1} \leftarrow P_1({\bf S}|{\bf D},C^i_{\ell})\, . 
\label{s_sample} 
\end{equation} 
}
\item{Now draw a sample of $C^{i+1}_{\ell}$ from the conditional density of $C_{\ell}$ given both, 
${\bf D}$ and ${\bf S}^{i+1}$, which was obtained in the first step above.  In symbols, 
\begin{equation}
C^{i+1}_{\ell} \leftarrow P_2(C_{\ell}|{\bf D},{\bf S}^i)\, .  
\label{cl_sample}
\end{equation} 
At this stage one has a pair of samples ${\bf S}^{i+1}, C^{i+1}_{\ell}$.
}
\item{Repeat above two basic steps for $i=1$ to $\mathcal N$, where $\mathcal N$ is a large number, 
 by replacing first $C^i_{\ell}$ by $C^{i+1}_{\ell}$ in step 1 above, and then
replacing ${\bf S}^{i+1}$ in Eqn.~\ref{cl_sample} by  the one  obtained in Eqn.~\ref{s_sample}.}
\end{enumerate}
After some initial pair of samples of signal and theoretical power spectrum are discarded (i.e., after 
initial burn-in period has completed)  they represent the desired samples drawn from the posterior density, $P({\bf S}, C_{\ell}| {\bf D})$, 
 under consideration.

\subsection{Density of pure CMB signal} 
 
The probability density function of pure CMB signal ${\bf S}$ given a theoretical CMB angular power 
spectrum $C_{\ell}$ is given by, 
\begin{eqnarray}
P_3({\bf S}| C_{\ell}) = \frac{1}{\sqrt{2^r\lambda_1\lambda_2...\lambda_r}}\exp{\left[-\frac{1}{2}{\bf S}^T {\bf C}^{\dagger} {\bf S}\right]}\, , 
\label{pdf_pure_s}
\end{eqnarray} 
where ${\bf C}$ denotes the $N \times N$ pixel-pixel CMB covariance matrix of ${\bf S}$ at the chosen beam and pixel 
resolution. As discussed in~\cite{Saha2017}, and as is the case in current article, rank $r$ of ${\bf C}$ is 
less than its size $N$, implying ${\bf C}$ is a singular 
matrix. ${\bf C }^{\dagger}$, therefore, represents the Moore-Penrose generalized inverse~\citep{Moore1920, Penrose1955} of ${\bf C}$.  The 
element of ${\bf C}$  can be computed from the knowledge of the CMB theoretical angular power spectrum  
using Eqn. 6 of~\cite{Saha2017},  by assuming that 
CMB map is  statistically isotropic.  The set of $\lambda_k, k \in \{1, 2, ..., r\}$ in the denominator of Eqn.~\ref{pdf_pure_s} represent 
the non-zero eigen values of  ${\bf C}$. 

\begin{figure*}
\includegraphics[scale=0.6]{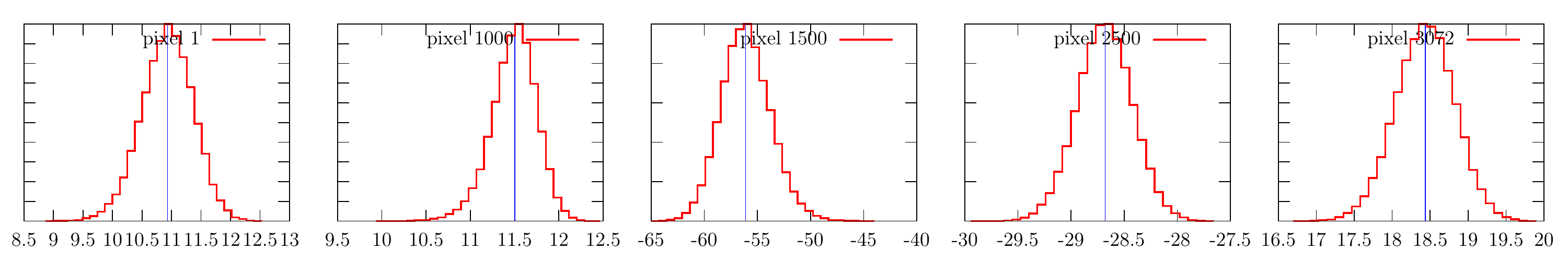}
\caption{The normalized probability density of CMB pixel temperatures for some selected pixels are shown in red. 
The normalization for each density is such that the peak corresponds to a value of unity. The horizontal 
axes represent pixel temperatures in the  unit of $\mu K$ (thermodynamic). The positions of mean temperatures are shown by the 
blue vertical lines. }
\label{hist_map}
\end{figure*}

\subsection{Drawing Samples of ${\bf S}$}
How do we draw samples of ${\bf S}$ given ${\bf D}$ and $C_{\ell}$? We must  do this without knowing or sampling the foreground
components, to keep our method foreground model independent. This will be possible if we could somehow remove 
all foregrounds without using their model, given ${\bf D}$ and $C_{\ell}$. The cleaned map obtained by using the global ILC method described 
in~\cite{Saha2017}  can be used  exactly for this purpose, if we assume that, the detector noise is  negligible  
and one has  sufficient number of input frequency maps to remove all foreground components, as discussed in the 
current section. Let us consider the cleaned map,
 ${\bf Y}$, obtained by using linear combination of $n$ frequency maps $\{{\bf X}_i\}$,    
\begin{equation}
{\bf Y} = \sum_{i=1}^{n} w_{i}{\bf X}_{i} \, ,
\label{cmap}
\end{equation}
where $w_{i}$ represents the weight corresponding to the $i^{th}$ input frequency map. Clearly, we can neglect 
any detector noise contribution in ${\bf Y}$ since ${\bf X}_i$ themselves are assumed to contain negligible 
detector noise (e.g., see Eqn.~\ref{cmb_fg}).
Since CMB follows blackbody 
distribution, to preserve the CMB signal in the cleaned map,  the weights for all frequency maps are constrained 
to add to unity, i.e, $w_1 + w_2 + w_3 + ... + w_n = 1$. Minimizing the CMB covariance weighted variance 
 $\sigma^2 = {\bf Y}^T {\bf C}^{\dagger} {\bf Y}$   of the  cleaned map ${\bf Y}$, subject to the above constraint 
on weights, as in~\cite{Saha2017}, 
we obtain, 
\begin{eqnarray}
{\bf W} = \frac{{\hat{\bf A}}^{\dagger}{\bf e}}{{\bf e}^T{\hat{\bf A}}^{\dagger}{\bf e}} \, , 
\label{weight}
\end{eqnarray} 
where ${\bf W}$ is an $n\times 1$ column vector with the $i^{th}$ element given by the weight factor $w_i$. ${\bf e}$ denotes 
an $n\times 1$ column vector with all entries equal to unity, representing the  frequency shape vector of the 
CMB component. Finally, the $(i,j)$ element of matrix ${\hat {\bf A}}$ can be computed in pixel-domain following,   
\begin{eqnarray}
\hat A_{ij} = {\bf X}_{i}^{T}{\bf C}^{\dagger}{\bf X}_{j}\, . 
\label{Aij} 
\end{eqnarray}
However, computing above in pixel-space is a numerically expensive process.  One can considerably simply 
Eqn.~\ref{Aij} in harmonic space. As shown in Appendix~\ref{A}, Eqn.~\ref{Aij} can conveniently be expressed 
in the multipole space following, 
\begin{eqnarray}
\hat A_{ij} = \sum_{\ell=2}^{\ell_{max}}\left(2\ell+1\right)\frac{\hat {\sigma}^{ij}_{\ell}}{C^{\prime}_{\ell}}\, ,
\label{Aij1}
\end{eqnarray}  
where, $C^{\prime}_{\ell}$ represents the beam and pixel smoothed CMB theoretical power spectrum, 
$C^{\prime}_{\ell} = C_{\ell} B^2_{\ell} P^2_{\ell}$, $B_{\ell}$ and  $P_{\ell}$ being the 
beam and pixel window functions respectively, and  $\hat {\sigma}^{ij}_{\ell}$ represents the 
cross angular power spectrum between frequency maps 
${\bf X}_i$ and ${\bf X}_j$~\footnote{ Eqn.~\ref{Aij1} becomes very useful when $\hat A_{ij}$ 
needs to be calculated repeatedly, such as, in the case of a Markhov chain.}.

 We assume that, there are  $n_f$ different foreground components, each with a constant
spectral index all over the sky~\footnote{The assumption of constant spectral index
of a component all over the sky is not necessarily a loss of generality, since as proposed
by~\cite{Francois1999} a foreground component with varying spectral index can be modeled in terms
more than one components each having different but constant spectral indices all over the sky.
Also see~\cite{Saha2016} for implementation of this concept using simulated observations of CMB
Stokes Q parameter. In our case, $n_f$ represents total number of all such components.}. 
We denote  the shape vector of $k^{th}$ foreground component by ${\bf f}_k$ (with $k \in \{1,2,..., n_f\}$)
, each one of which  is an $n \times 1$
column vector. Using Eqns.~\ref{cmb_fg} and \ref{weight} in Eqn.~\ref{cmap} we obtain,
\begin{eqnarray}
{\bf Y} = {\bf S} + \left[{\bf W}^T\sum_{k=1}^{n_f}{\bf f}_k \right]{\bf F}^0_k\, , 
\label{cmap1}
\end{eqnarray}
where ${\bf F}^0_k$ is an  $N \times 1$ column vector representing an appropriately 
chosen template for the $k$ foreground component. Eqn.~\ref{cmap1} shows that the 
cleaned maps contains the pure CMB signal plus some foreground residual given by the 
second term.  To find these residuals, introducing matrix notation, we first write Eqn.~\ref{Aij1}
as
\begin{eqnarray}
{\hat {\bf A}} = \sum_{\ell=2}^{\ell_{max}}\frac{\left(2\ell+1\right)}{C^{\prime}_{\ell}}{\hat {\boldsymbol \Sigma}}_{\ell}\, ,
\label{amatrix} 
\end{eqnarray}   
where $n \times n$ data covariance matrix, $\hat {\boldsymbol \Sigma}_{\ell}$, in the harmonic space can be written in terms of CMB angular 
power spectrum, $\hat C_{\ell}$, of the particular random realization under consideration and 
the foreground covariance matrix ${\bf C}^f_{\ell}$, as, 
\begin{eqnarray}
\hat {\boldsymbol \Sigma}_{\ell} = \left[{\bf ee}^T\hat C_{\ell}  + {\bf C}^f_{\ell}\right]B^2_{\ell}P^2_{\ell} \, .
\label{Sigmal}
\end{eqnarray} 
Using Eqn.~\ref{Sigmal} and~\ref{amatrix} in Eqn.~\ref{weight} and following a procedure similar to~\cite{Saha2016}
we  obtain, 
\begin{eqnarray}
{\bf W} = \frac{\left({\bf I} - { { {\bf C}_f}}{ {{\bf  C}_f}}^{\dagger}\right){\bf{e}}}
{{\bf e}^T\left({\bf I} - { {{\bf C}_f}}{ { {\bf C}_f}}^{\dagger}\right){\bf{e}}}\, ,
\end{eqnarray}
where ${\bf I}$ denotes the $n \times n$ identity matrix and 
\begin{eqnarray} 
  {\bf C}_f = \sum_{\ell=2}^{\ell_{max}}\frac{\left(2\ell+1\right)}{C_{\ell}}{\bf C}^f_{\ell}\, ,   
\end{eqnarray} 
The product $ {\bf C}_f{\bf C}^{ \dagger}_f$ represents the projector on the column space, $\mathcal C({\bf C}^f)$,
of ${\bf C}_f$. If we assume that  $n_f < n$, which may be achieved by using sufficiently large 
number of input frequency maps,
the null space of ${\bf C}_f$ is an non-empty set and ${\bf I} -  {\bf C}_f{ \bf  C}^{\dagger}_f$
is a projector on this null space. Since, the shape vector, ${\bf f}_k$ 
 of each foreground components with constant spectral indices completely lies on $\mathcal C({\bf C}_f)$
we must have, ${\bf W}^T{\bf f}_k = 0$, for all $k$. Therefore, from Eqn.~\ref{cmap1} one finds that the foreground contamination in the final cleaned map 
at each pixel due to all foreground components disappears. Hence ${\bf Y} = {\bf S}$. 

Based upon preceding discussions, to sample ${\bf S}$  from $P_1({\bf S}|C_{\ell}, {\bf D})$ we use Eqn.~\ref{cmap}. 
The cleaned map in this case has the  probability density as given by Eqn.~\ref{pdf_pure_s} with the 
same covariance structure mentioned therein.  

\subsection{Drawing Samples of $C_{\ell}$}
\label{Cl_sample} 
As discussed in Appendix~\ref{B} the conditional density $P_2(C_{\ell}|{\bf S},{\bf D})$ can be written as, 
\begin{eqnarray}
P_2\left( C_{\ell}|\hat C_{\ell}\right) \propto \left(\frac{1}{C_{\ell}}\right)^{\left(2\ell+1\right)/2}
\exp\left[-\frac{\hat C_{\ell}\left(2\ell+1\right)}{2C_{\ell}}\right] \, , 
\label{pdf_cl1}
\end{eqnarray}
where the variable $z = {\hat C_{\ell}\left(2\ell+1\right)}/{C_{\ell}}$ follows a $\chi^2$ distribution 
of $2\ell-1$ degree of freedom. To draw samples of $C_{\ell}$ from Eqn.~\ref{pdf_cl1} we first draw $z$
from  the $\chi^2$ distribution of $2\ell-1$ degrees of freedom, which is achieved by drawing $2\ell-1$
independent standard normal deviates and forming the sum of their squares. Given the value of $\hat C_{\ell}$
estimated from the map, we then find $C_{\ell}$ following  $C_{\ell} =  {\hat C_{\ell}\left(2\ell+1\right)}/z$.

\section{Methodology}
\label{Method} 
\begin{figure}
\includegraphics[scale=0.3]{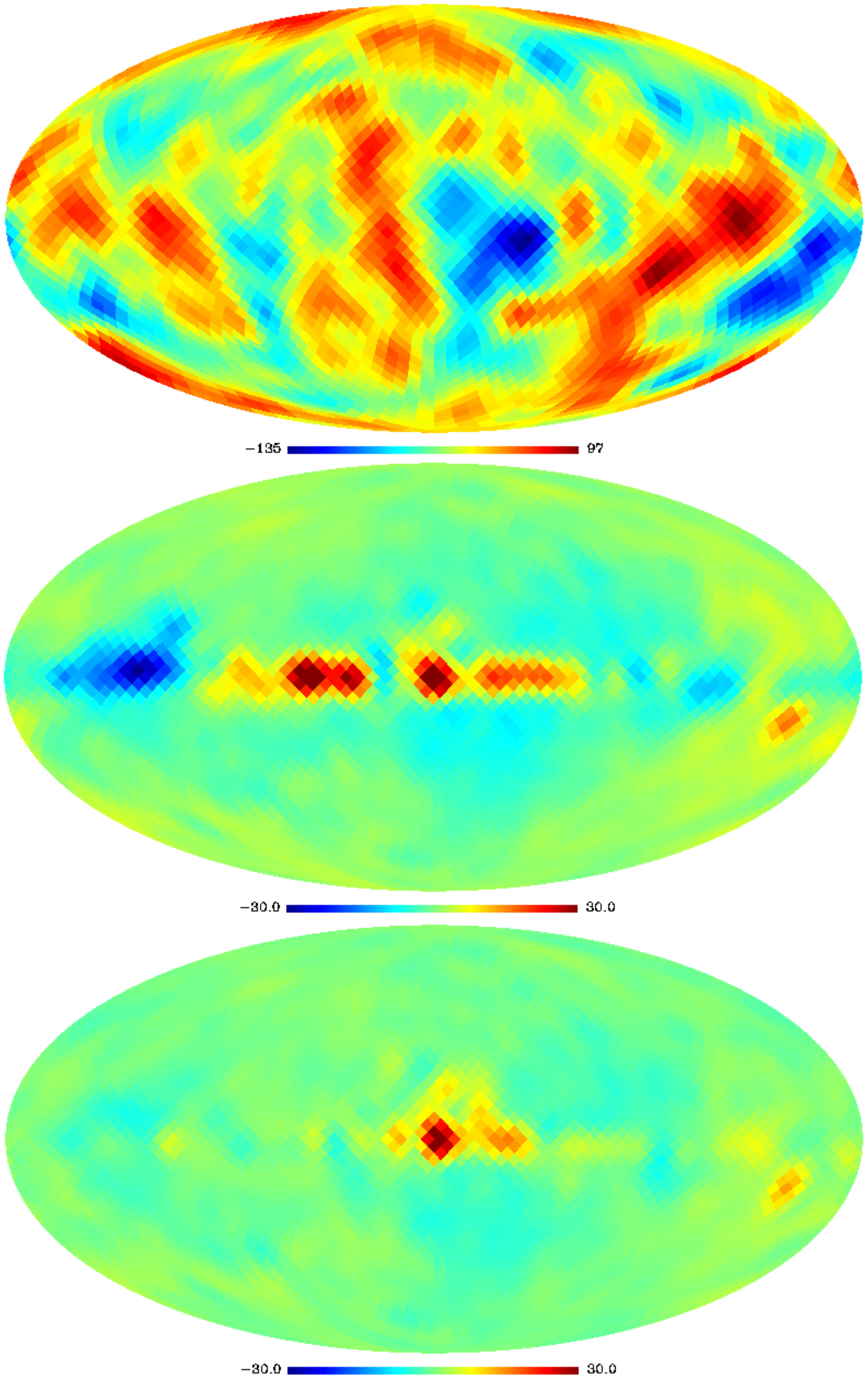}
\caption{Top panel shows the best-fit CMB map obtained by our method. The middle and bottom panels show the difference 
of our map from the Commander and  NILC cleaned maps respectively. There is a noticeable similarity between  the best-fit 
and NILC cleaned maps as seen from the bottom panel. }
\label{best_fit_cmb}
\end{figure}

\begin{figure}
\includegraphics[scale=0.3]{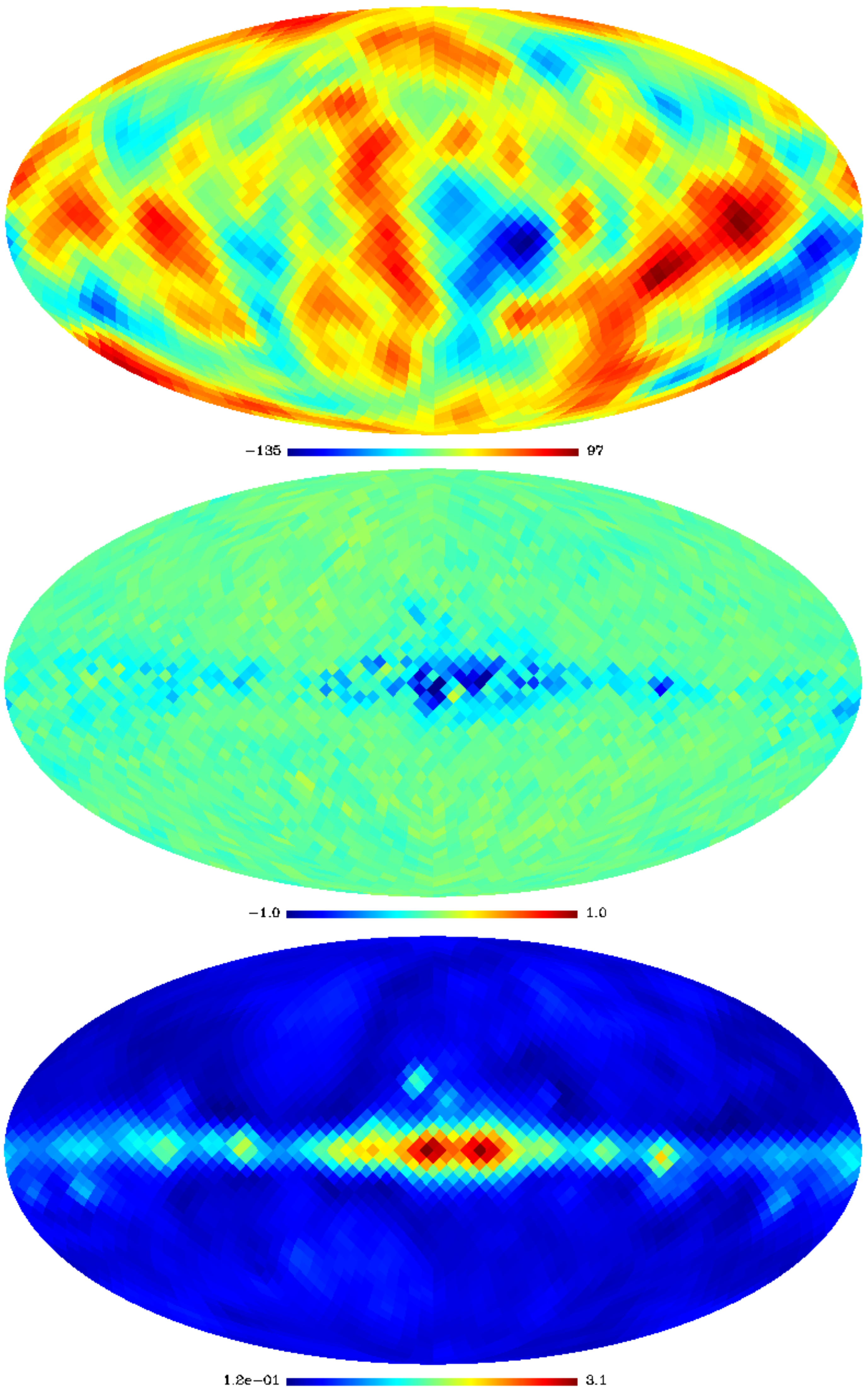}
\caption{Top panel shows the mean CMB map estimated by using all the cleaned maps obtained
from the Gibbs samples. The mean map matches very well with the best-fit map shown in the top panel of Fig.~\ref{best_fit_cmb}.
The middle panel show the difference between the best fit and the mean CMB maps. The bottom panel shows
the standard deviation map obtained  by using all the cleaned maps.  }
\label{mean_diff}
\end{figure}

\begin{figure*}
\includegraphics[scale=0.7]{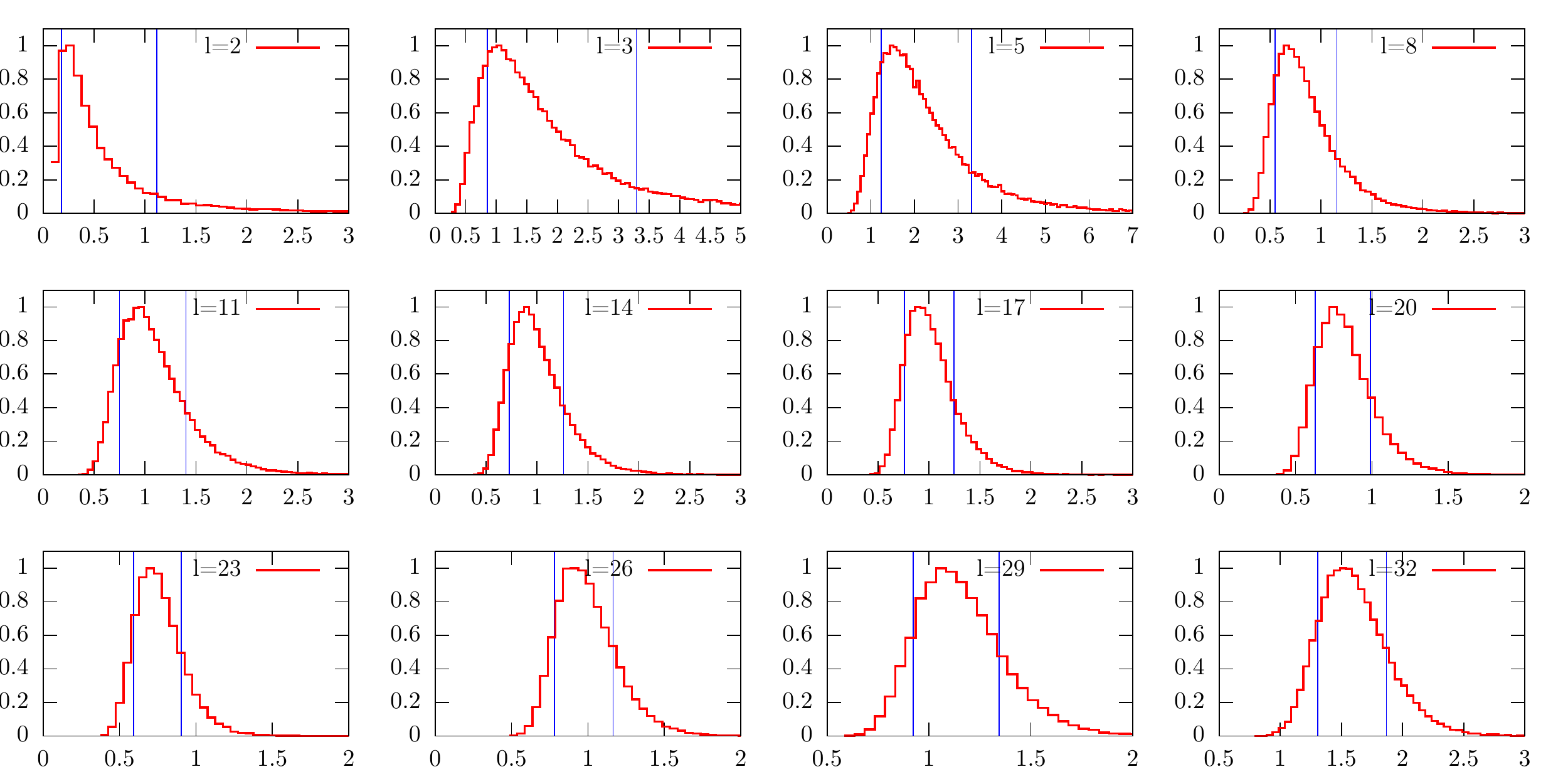}
\caption{Normalized densities of the CMB theoretical angular power spectrum obtained by Gibbs sampling
for different multipoles. The horizontal axis for each sub plot represents  $\ell\left(\ell +1\right)C_{\ell}/(2\pi)$ in the unit 
$1000$ $\mu K^2$. The region within the two vertical lines represent $1-\sigma$ confidence interval for the 
theoretical angular power spectrum. }
\label{hist_cls}
\end{figure*}

\begin{figure*}
\includegraphics[scale=0.7]{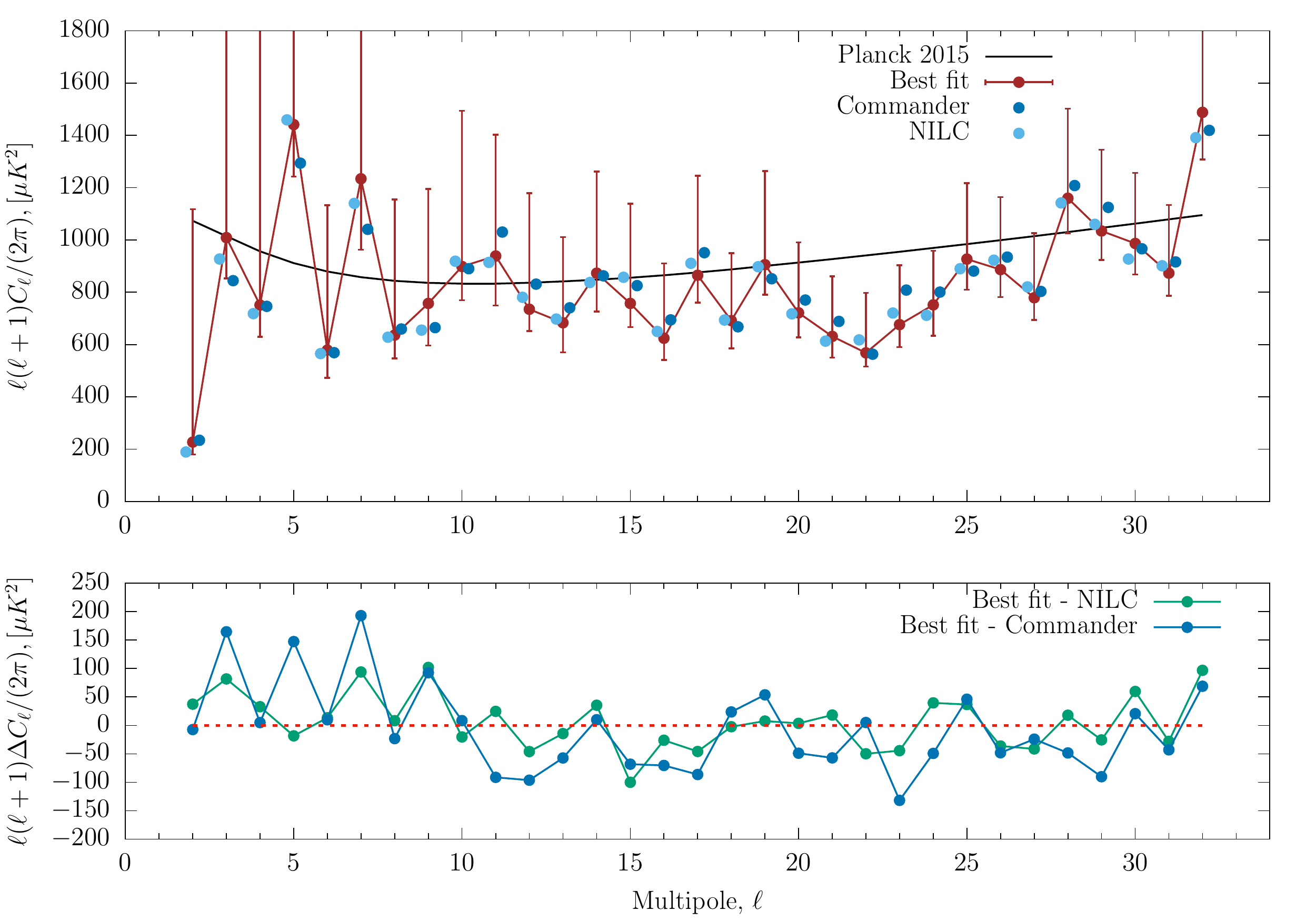}
\caption{Top panel shows the best-fit CMB theoretical angular power spectrum  along with the 
asymmetric error bars indicating $68.27\%$ confidence intervals obtained 
from the Gibbs samples in brown line. The sky blue points represent the angular power spectrum estimated 
from NILC CMB map. The deep  blue points represent the same estimated from the Commander CMB maps. 
(For visual purpose, both these spectra are shifted along the horizontal axis slightly from the actual 
positions of the integer multipoles.) The black line shows the Planck 2015 theoretical power spectrum 
as a guide to eye. The bottom panel shows a zoomed in version  of the  differences  of Commander and NILC 
angular power spectra respectively from the best-fit  angular power spectrum of top 
panel.    }   
\label{cl}
\end{figure*}

We use WMAP nine-year $10$ difference assembly (DA) maps and seven Planck 2015 
maps,  three of the later are at  LFI frequencies ($30$, $40$ and $70$ GHz) and the rest 
at four HFI frequencies ($100, 143, 217$
and $353$ GHz). The processing of input maps remains identical to~\cite{Saha2017} and results in
a total of $12$ input maps,  five at WMAP and seven at Planck frequencies.  We  note that, 
since we are interested in analysis over large angular scales  of the sky, where detector
noise can be ignored,  we chose 
a low pixel resolution defined by HEALPix~\footnote{Hierarchical Equal Area isoLatitude 
Pixellization of the sky, a freely available software package for analysis of CMB 
maps, e.g., see~\cite{Gorski2005}.} parameter $N_{side}$ = $16$ and a Gaussian beam smoothing 
of $9^\circ$ FWHM for each input map. We remove both monopole and dipole from all the full sky 
input maps before the analysis. To sample the posterior density $P({\bf S}, C_{\ell}|{\bf D})$
we simulate a total of $10$ Gibbs chains, each containing $5000$ joint samples of cleaned maps and 
theoretical power spectrum, following the three sampling steps described in Section~\ref{formalism}.    
At any given chain and at any given iteration, to sample ${\bf S}$ we use Eqn.~\ref{cmap}, where the weights are described by the vector ${\bf W}$
defined by Eqn.~\ref{weight}. The elements of matrix ${\bf A}$ that appears in~\ref{weight} 
 are computed following Eqn.~\ref{Aij1} using   the last sampled $C_{\ell}$ values.  After sampling ${\bf S}$
we estimate its full sky power spectrum , $\hat C_{\ell}$, which we use to obtain a new sample of $C_{\ell}$
using the method described in Section~\ref{Cl_sample}.  We emphasize that both $\hat C_{\ell}$ and $C_{\ell}$ 
contain beam and pixel smoothing effects absorbed in them. We note in passing, that the weights as given by 
Eqn.~\ref{weight} are however insensitive to such smoothing since in Eqn.~\ref{Aij1}, both $C_{\ell}$ and $\hat{\sigma}^{ij}_{\ell}$ contain
same smoothing effects. The initial choice of $C_{\ell}$ for each  chain is made by  drawing  them uniformly within $\pm 3\Delta C_{\ell}$
around the Planck best-fit theoretical power spectrum~\citep{PlanckCosmoParam2016}, where $\Delta C_{\ell}$ denotes error due to cosmic 
variance alone.  

All $10$ chains generates a total of $50000$ joint samples of cleaned map and theoretical CMB 
power spectrum. The burn in phase in each chain is very brief. Visually, this phase does not 
appear to contain more than a few  Gibbs iterations. We, however, remove $50$ initial 
Gibbs iterations from each chain as a conservative estimate of burn-in period. After the burn-in
rejection we have a total of $49500$ samples from all chains.

\section{Results}
\label{Results}
\subsection{Cleaned Maps}
\label{cmaps}

Using all samples after burn in rejection we estimate the marginalized  probability density of CMB temperature at each
pixel given the observed data. The normalized probability densities obtained by division by the corresponding 
mode of the marginalized density function for some selected pixels over the sky are shown in Fig.~\ref{hist_map}. These density functions are approximately
symmetric with some visible asymmetry near the tails. The positions of mean temperatures are shown by the 
blue vertical lines for each pixel of this plot. We estimate the best-fit CMB cleaned map by taking the
pixel temperatures corresponding to the location of mode of density for each pixel at $N_{side} = 16$. 
We show the best-fit map at the top panel of Fig.~\ref{best_fit_cmb}. We compare  the
best-fit map with other CMB cleaned maps which are obtained by using different methods by other science
groups. We show the differences of best fit map from Planck Commander and NILC (needlet space ILC)
cleaned maps~\citep{Planck2016_CMB} respectively
at the  middle and bottom panels of the same figure at $9^{\circ}$ Gaussian beam resolution. Clearly, our best-fit map matches well with the Commander 
cleaned map with some minor differences along the galactic plane, which is expected to contain 
some foreground residuals in any foreground removal method. It is worth to emphasize  the striking similarity 
between  the best-fit  and NILC CMB map. Interestingly, the best-fit map contains somewhat lower pixel temperatures  
at isolated locations along the galactic plane than the Commander or NILC map.

From Fig.~\ref{hist_map} we see that the mean and marginalized posterior maximum of CMB 
temperature at different pixels agree closely with each other. We estimate the mean CMB map
using all $49500$ samples for each pixel and show this at the top panel of Fig.~\ref{mean_diff}.
The mean map matches very well with the best-fit CMB map shown in top panel of the Fig.~\ref{best_fit_cmb}.
We have plotted the difference between the best fit and mean map in the bottom panel of Fig.~\ref{mean_diff}.
Both the maps agree with each other within an absolute difference of $1\mu K$. In order to quantify the 
reconstruction error in the cleaned CMB map obtained after each iteration of Gibbs sampling,
we generate a standard deviation map using all $49500$ cleaned maps. We show this map at the bottom panel 
of Fig.~\ref{mean_diff}. From this panel we see that the reconstruction error is very small all over the 
sky. The maximum reconstruction error is visible along the galactic plane and towards the center of 
our galaxy where the input frequency maps contain strong foreground contaminations. 
 
\subsection{Angular Power Spectrum}
\label{theorycl}

We estimate the marginalized probability density of CMB theoretical angular power spectrum and show the 
results in Fig.~\ref{hist_cls} for different multipoles. Like the density functions of the pixel temperatures 
as discussed in Section~\ref{cmaps}, we normalize these densities to a value of unity at their peaks. 
The horizontal axis of each plot of this figure represents $\ell\left(\ell +1\right)C_{\ell}/(2\pi)$ in the unit 
of $1000$ $\mu K^2$. The density functions show long asymmetric tails for low multipoles (e.g., $\ell = 2, 3, 5$). For large
 multipoles ($\ell \ge 20$) the asymmetry of the densities become gradually reduced. The region within the two vertical lines in each plot show 
$1-\sigma$ ($68.27\%$) confidence interval for the CMB theoretical power spectrum for the corresponding multipole.      

In top panel of Fig.~\ref{cl} we show the best-fit theoretical CMB angular power spectrum in brown color, 
defined by the positions of peaks of marginalized angular power spectrum density functions (e.g., 
Fig~\ref{hist_cls}). The asymmetric error bars at each $\ell$ 
show the $1-\sigma$ confidence interval for the theoretical angular power spectrum. The black line 
shows the theoretical power spectrum consistent with Planck 2015 results~\citep{PlanckCosmoParam2016}.
The best-fit theoretical angular power spectrum agree well with the spectra estimated from Commander and 
NILC cleaned maps, which are shown by green and blue points respectively.
In the bottom 
panel of  Fig.~\ref{cl}  we show the difference of the best fit and Commander (or NILC) power spectrum. 
Our best fit theoretical power spectrum agrees very well with the spectrum estimated from the NILC 
CMB map.

\section{Convergence Tests}
\label{convergence}
Each of the Gibbs sampling chains for the estimation of CMB signal and its theoretical 
angular power spectrum joint posterior consists of $4950$ sampling after rejection of burn-in phase. 
A diagnosis is necessary to be certain that these samples have converged to the actual targeted 
CMB  posterior - a condition when satisfied inference drawn about any parameter by using the chains,
does not depend upon the initial point where the chain starts.~\cite{GR1992} propose that lack 
of any such convergence is better diagnosed if we simulate a  
set of `parallel' chains than a single chain. Using all the $10$ Gibbs chains we, therefore, 
check for convergence  by using the Gelman-Rubin statistic~\cite{GR1992}. Detailed description of 
the statistic is given in~\cite{GR1992,Brooks1998}, however, for completeness we define the
statistic below.

Let us assume that we have generated  $M$  number of different chains and let $L$ be the 
number of steps  in each chain after rejection of samples during the burn-in period~\footnote{$L$ 
can be different for different chains, however, if $L$ is same for all chains simplifies 
calculations.}.  For a model parameter $\theta$, let us assume that the sample posterior 
mean is given by $\bar{\theta}_{m}$ for $m^{th}$ chain using all  $L$ samples.  Let corresponding 
sample posterior variance is $\bar{\sigma}^2_{m}$. Then 
the between-chain ($B/L$) and within-chain variances ($W$) are  respectively given by, 
\begin{eqnarray}
B &= &\frac{L}{M - 1}\sum_{m=1}^{M}\big(\bar{\theta}_{m} - \bar{\theta}  \big)^2 \\
W &= &\frac{1}{M}\sum_{m=1}^{M}\bar{\sigma}^2_{m} \, , 
\end{eqnarray} 
where $\bar{\theta}$ is the overall posterior mean of the samples estimated from  all $M$ chains and is
given by $\bar{\theta} = \frac{1}{M}\sum_{m=1}^{M}\theta_{m}$.  We define the pooled posterior 
variance following,
\begin{equation}
\hat{V} = \frac{L - 1}{L}W + \frac{M + 1}{ML}B\, ,
\end{equation}
which can be used  to compute the Gelman-Rubin statistic $R$ as follows, 
\begin{equation}
R= \sqrt{\frac{\hat V}{W}}\, . 
\end{equation}
Following~\cite{GR1992, Brooks1998} a value of $R$ close to unity implies that each 
of $M$ Gibbs chains have converged to the target posterior density.      

\begin{figure}
\includegraphics[scale=0.68]{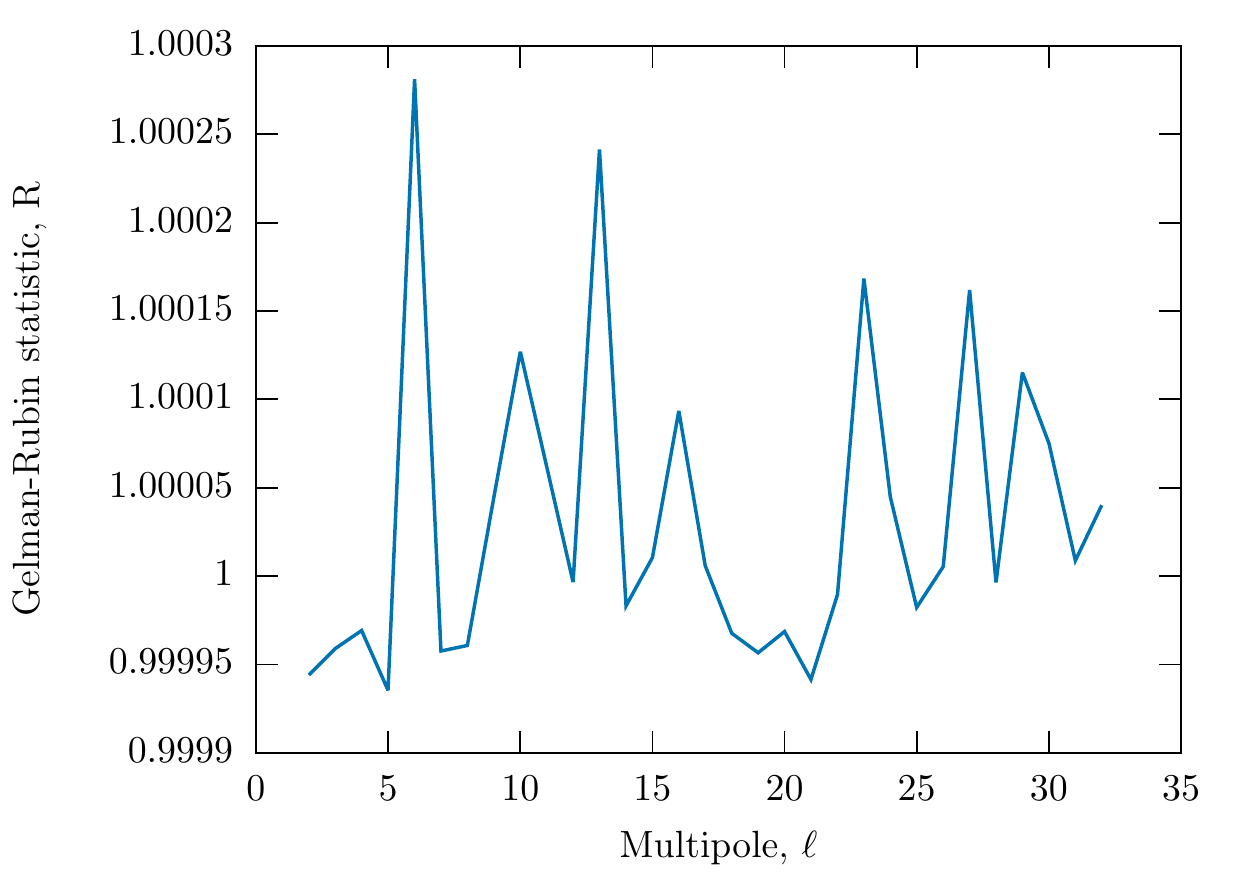}
\includegraphics[scale=0.32]{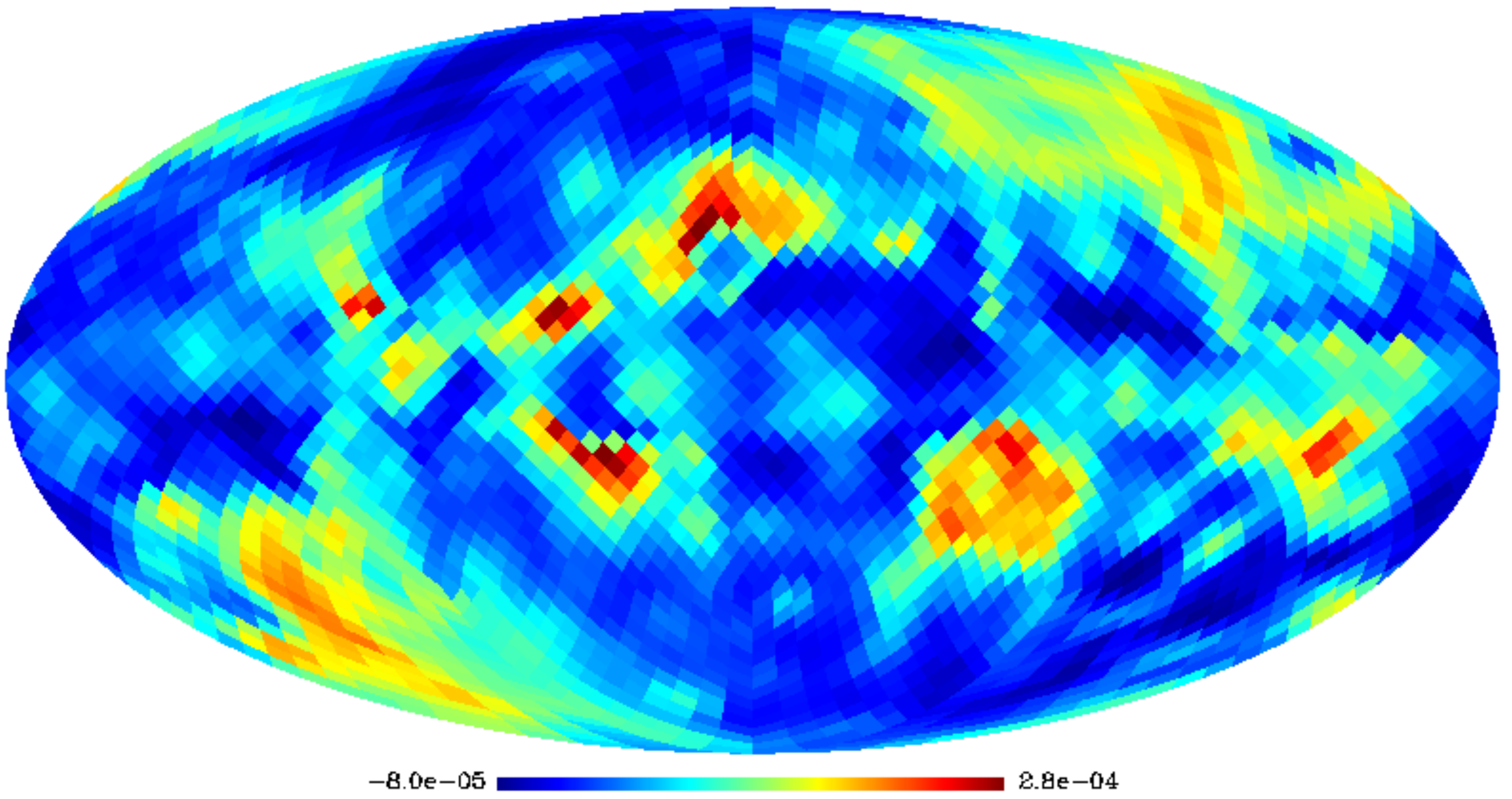}
\caption{Top panel shows Gelman-Rubin statistic, $R$ estimated for all the multipoles 
of this work. The bottom panel shows map of $R-1$ for all pixels. Close values of $R$ to unity 
in both cases indicate convergence is achieved in all the Gibbs chains.}
\label{GR}
\end{figure}

We have plotted the Gelman-Rubin statistic, $R$, for the theoretical angular power spectrum
samples for the multipole range $2 \le \ell \le 32$  in top panel of Fig.~\ref{GR}. The value 
of $R$ lies well within $0.9999$ and $1.0003$ implying convergence. In the bottom panel of Fig.~\ref{GR}   
we show the map of $R-1$ all over the sky. $R$ lies with in $0.99992$ and $1.0002$ for all the pixels
implying again convergence of the Gibbs chains.   

\section{Monte Carlo Simulations}
\begin{figure}
\includegraphics[scale=0.32]{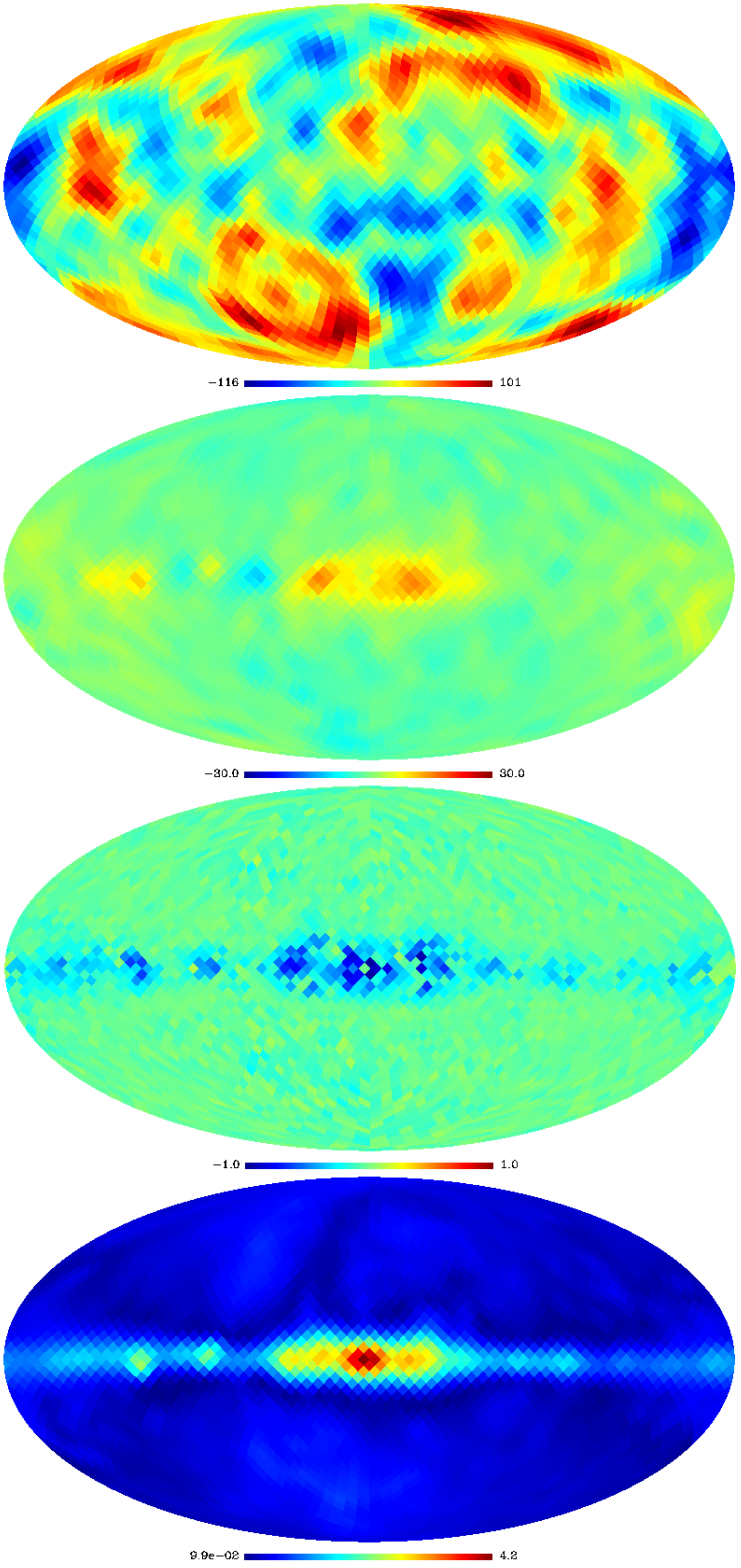}
\caption{Top panel shows the best-fit cleaned CMB map obtained from Monte Carlo simulations. The second panel
shows the difference between the best-fit and the input CMB map used in the simulation. The third panel shows the
difference between the best-fit and mean CMB map. The last panel shows the standard deviation map. }
\label{sim_maps}
\end{figure}

\label{MC}
We generate a set of input maps at the $12$ different WMAP and Planck frequencies at 
a Gaussian beam resolution $9^{\circ}$ and $N_{side} = 16$ 
following the same procedure as described in~\cite{Saha2017}. We do not reproduce the 
methodology in the article and refer to the above article for a description about the 
input frequency maps.  We note that, in the current work, we need to simulate only
one random realization of the $12$ input frequency maps. The random CMB realization used 
in the input frequency maps is generated using the CMB theoretical angular power spectrum 
consistent with Planck 2015 results.  As is the case for our analysis on the Planck and 
WMAP observations, we remove both monopole and 
dipole from all the simulated input maps before sampling from the posterior density $P({\bf S}, C_{\ell}| {\bf D})$.    
We simulate a total of $10$ Gibbs chains.  To draw the first cleaned map sample we  initialize the 
theoretical CMB power spectrum  uniformly  within $\pm 3\Delta_{\ell}$ of the true theoretical spectrum,
where $\Delta_{\ell}$ represents cosmic variance induced error. As in the case of analysis of 
WMAP and Plank observed maps, for simulations also we find that the burn-in period ends very rapidly. 
In particular, from the trace plots of pixel temperature of the cleaned maps we see that burn-in 
phase completes within a few samples. As a conservative approach, however, we reject initial $50$
samples from each Gibbs chain. After, burn-in rejection we have a set of $4950$ joint samples of
cleaned map and theoretical angular power spectrum from each Gibbs chain.      

Using all cleaned map samples from all chains after burn-in rejection we form a marginalized density 
of the CMB temperature at each $N_{side} = 16$ pixel. A CMB map formed from the pixel 
temperatures corresponding to the modes of these density functions define the best fit 
CMB cleaned map obtained from the simulation. We show the best-fit cleaned map in top panel 
of Fig.~\ref{sim_maps}. The difference of the best-fit and input CMB realization
is shown in the second panel of the same figure. Clearly, the best-fit CMB map matches very well
with the CMB realization used in the simulation. The maximum difference ($15.15$ $\mu K$) 
between the two maps is observed along the galactic plane. This shows that our method
removes foreground reliably. The third panel of Fig.~\ref{sim_maps} shows the difference 
of best-fit and mean CMB maps obtained from all Gibbs samples. Both these maps agree 
very well with each other. The last panel of Fig.~\ref{sim_maps} shows the  standard deviation map
computed from the Gibbs samples. The maximum error of $4.2$ $\mu K$ is observed at the galactic center.
In summary, using the Monte-Carlo simulations of our method,  we see that the best-fit and 
mean CMB maps agree very well with the input CMB map indicating a reliable  foreground 
minimization can be achieved by our method.

\begin{figure}
\includegraphics[scale=0.7]{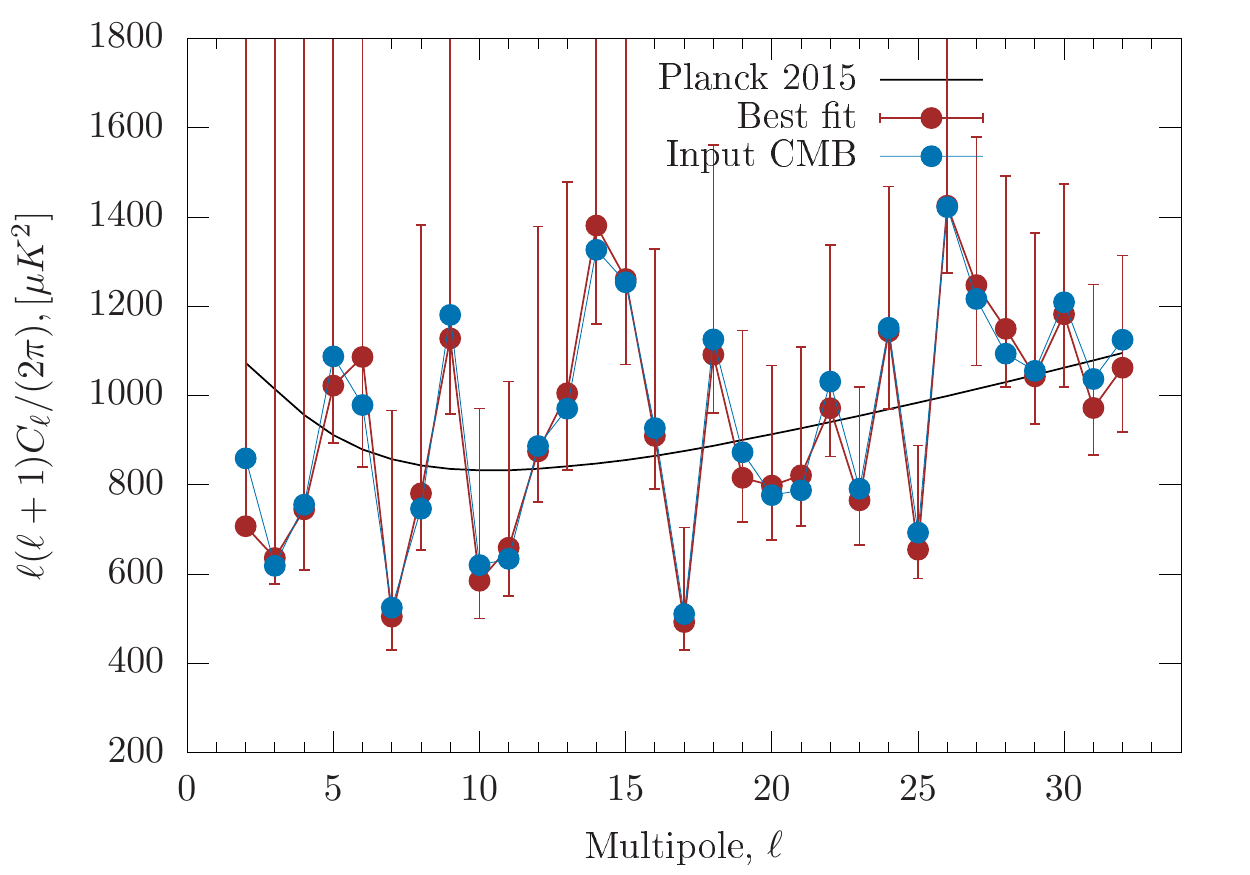}
\caption{Figure showing the best-fit estimate of theoretical CMB angular power spectrum (in brown)
with $68.27\%$ confidence intervals for different multipoles along with the CMB angular power 
spectrum (in blue) estimated from the specific CMB realization used in the Monte Carlo simulations. 
The black line indicates the theoretical angular power spectrum from which the specific CMB realization 
under consideration is generated.}
\label{cl_sim}
\end{figure}

We show the best-fit estimate of underlying CMB theoretical angular power 
spectrum from Monte Carlo simulations in Fig.~\ref{cl_sim} in 
brown line. The asymmetric  error-limits shown on this power spectrum 
indicate the $68.27\%$ confidence intervals. The angular power spectrum 
for the input CMB map used in the simulation is shown in blue. The best-fit
estimate agrees nicely with the input angular power spectrum. The black
line of this figure represents the underlying  theoretical angular power
spectrum that is used to generate the input CMB realization for the simulation. 

We test the Gibbs sequences obtained from the simulations for convergence 
as in Section~\ref{convergence} using the Gelman-Rubin statistic, $R$. The maximum
and minimum values of $R$ for the sampled CMB maps over all pixels 
are respectively, $1.00017$ and $0.999924$. The corresponding values for the sampled 
angular power spectra are respectively,   $1.00027$  and $0.999935$. Such values of $R$ close to unity 
indicate convergence of the Gibbs sequences in Monte Carlo simulations.

\section{Discussions and Conclusions}
\label{Discussion}
In this article, we have presented a new method to estimate the CMB posterior density over the 
large angular scales of the sky, given the Planck and WMAP observations by using 
a global ILC method~\citep{Saha2017} and Gibbs sampling~\citep{GR1992} as the basic tools.  Our main results are joint estimates of best-fit 
CMB signal and its theoretical angular power spectrum along with the appropriate 
confidence intervals which can be directly used for cosmological parameter estimation. 
Therefore, our work, for the first time effectively extends the ILC method for such purposes. 
 We sample the CMB signal at each Gibbs iteration conditioning on a set of CMB theoretical 
angular power spectrum obtained in the previous Gibbs iteration. The CMB reconstruction step
is independent on any explicit model of foreground components - which is a characteristic of 
the usual ILC method. However, considering the sampling of both CMB signal and its theoretical angular 
power spectrum, the new method extends the model independent nature of CMB reconstruction of
ILC method to  the entire posterior 
density estimation at low resolution. Thus our method serves as a complementary route 
to the CMB posterior estimation where detailed model of foregrounds are taken into account.
We have implemented the CMB reconstruction method in the harmonic space which reduces computational
time significantly unlike the pixel-space approach of~\cite{Saha2017}.      

There are some aspects of the method which one needs to address in 
future investigations. In the current work we have assumed that the detector noise can be 
completely ignored which is a valid assumption on the large angular scales for experiments like 
WMAP and Planck. A general framework will be to formalize the method in the presence of detector noise.
In the presence of detector noise the blind foreground removal procedure will leave some foreground 
residuals on the cleaned maps. It would be interesting to see  whether these residuals can 
be taken care of using a foreground model independent manner. 

In the current method where detector noise is assumed to be negligible residual foregrounds 
will be present in the cleaned maps if effective number of foreground components $n_f$ 
present in the  input frequency maps become larger than or equal to number of input frequency maps ($n$)
available. For large angular scale analysis like the one of this paper, $n_f < n$ is a reasonable 
assumption outside the galactic plane. Along the plane, where the foreground spectral properties are 
expected show a larger variation than the outside plane, the effect of such residuals are mitigated by the 
smoothing of the input sky maps over the large angular scales. By performing detailed Monte Carlo 
simulations we see that the method leaves a small residual along the galactic plane. By comparing our cleaned map and angular power 
spectrum results with those obtained by other science groups we show that the level of such 
foreground residuals are small and of comparable magnitudes of those  present in CMB maps 
obtained by other methods.   

We thank an anonymous referee of an earlier publication by the authors~\citep{Saha2017} for suggestions
of integrating the work with Gibbs sampling.  Our work is based on observations obtained with Planck (http://www.esa.int/Planck), 
an ESA science mission with instruments and contributions directly funded by ESA Member States, 
NASA, and Canada.  We use publicly available HEALPix~\cite{Gorski2005} package  
(http://healpix.sourceforge.net) for some of the analysis of this work. 
We acknowledge the use of Planck Legacy Archive (PLA) and the Legacy Archive for Microwave Background 
Data Analysis (LAMBDA). LAMBDA is a part of the High Energy Astrophysics Science Archive Center (HEASARC). 
HEASARC/LAMBDA is supported by the Astrophysics Science Division at the NASA Goddard Space Flight
Center. 

\appendix

\section{A:  Elements of Matrix ${\bf A}$ in Harmonic Space}
\label{A} 
Let us assume that, $S(p)$ denotes a random simulation of a pixellized CMB map ($p$ denotes pixel index)
at some beam and pixel resolutions. 
Using the spherical harmonic decomposition $S(p) = \sum_{\ell,m} a_{\ell,m} Y_{\ell m}( p)$, the $(p,q)$ 
element, $C_{pq}$  of the pixel-pixel CMB covariance matrix, ${\bf C}$ can be written as
\begin{eqnarray}
C_{pq} = \left<S(p)S(q)\right> = \sum_{\ell m \ell'  m'} \left< a_{\ell m} a^*_{\ell'm'}\right> Y_{\ell m}(p)Y^*_{\ell'm'}(q)\, ,
\end{eqnarray}  
where $\left < ...\right>$ represents ensemble average and we assume that the beam and pixel smoothing effects 
are implicitly contained in spherical harmonic coefficients $a_{\ell m}$. Using statistical isotropy of CMB, namely, 
$ \left< a_{\ell m} a^*_{\ell'm'}\right>   = C^{\prime}_{\ell}\delta_{\ell \ell'}\delta_{mm'}$, where 
$C^{\prime}_{\ell} = C_{\ell}B^2_{\ell}P^2_{\ell}$, ($B_{\ell}$ and $P_{\ell}$ being respectively
beam and pixel window functions)  we obtain, 
\begin{eqnarray}
C_{pq} = \sum_{\ell m } C^{\prime}_{\ell} Y_{\ell m}(p)Y^*_{\ell m}(q)\, .  
\label{cpq_iso} 
\end{eqnarray}
In matrix notation we write Eqn.~\ref{cpq_iso} as 
\begin{eqnarray}
{\bf C} = \sum_{\ell m} C^{\prime}_{\ell} {\bf D}_{\ell m} \, ,
\end{eqnarray} 
where elements of matrix ${\bf D}_{\ell m}$ are given by $D^{pq}_{\ell m} = Y_{\ell m}(p)Y^*_{\ell'm'}(q)$. 
We can write, ${\bf D}_{\ell m} = {\bf Z}_{\ell m} {\bf Z}^c_{\ell m}$, where, ${\bf Z}_{\ell m}$ is an $N \times 1$ column
vector with elements $Z_{\ell m}(p) = Y_{\ell m}(p)$, and the superscript ${}^c$ represents the Hermitian 
conjugate.  
It is worth emphasizing that, the right hand side of above equation represents a linear combination of ${\bf D}_{\ell m}$ 
matrices with the scalar amplitudes given by $C^{\prime}_{\ell}$. Clearly, therefore,  
\begin{eqnarray}
{\bf C}^{\dagger} = \sum_{\ell m} {\bf D}^{\dagger}_{\ell m} C^{\prime \dagger}_{\ell} \, . 
\label{cdagger}
\end{eqnarray}
Using the definition of Moore-Penrose generalized inverse, ${\bf x}^{\dagger}$ of a vector ${\bf x}$, 
${\bf x}^{\dagger} = {\bf x}^c/||{\bf x}||^2$ where $||...||^2$ represent squared norm of the vector, one 
obtains,
\begin{eqnarray}
{\bf D}^{\dagger}_{\ell m} = \frac{{\bf Z}_{\ell m} {\bf Z}^c_{\ell m}}{||{\bf Z}_{\ell m}||^4}\, . 
\label{ddagger}
\end{eqnarray}   
Using orthogonality of spherical harmonics over discrete HEALPix pixels 
\begin{eqnarray}
\left[\sum_{p=1}^N Y_{\ell m}(p)  Y^*_{\ell m}(p)\right] \frac{4\pi}{N}  = 1 \, ,
\label{oc}
\end{eqnarray} 
it is easy to find $||{\bf Z}_{\ell m}||^4 = \left(N/(4\pi)\right)^2$.  Using this result and Eqn.~\ref{ddagger} in Eqn.~\ref{cdagger}
we obtain element wise,  
\begin{eqnarray}
C^{\dagger}_{pq} = \sum_{\ell m}\frac{1}{C^{\prime}_{\ell}}\frac{Y_{\ell m}(p) Y^*_{\ell m}(q)}{\left(\frac{N}{4\pi}\right)^2}\, ,  
\label{cdagger1} 
\end{eqnarray} 
where we have used $C^{\prime \dagger}_{\ell} = 1/C^{\prime}_{\ell}$. Expanding the input frequency maps ${\bf X}_i$ in spherical 
harmonic space, $X_i(p) = \sum_{\ell_1 m_1} a^i_{\ell_1, m_1} Y_{\ell_1 m_1}(p)$, using  Eqn.~\ref{cdagger1} and 
the orthogonality condition of spherical harmonics as mentioned in Eqn.~\ref{oc}, after some algebra,   we obtain, 
\begin{eqnarray}
\hat A_{ij} = {\bf X}^T_i {\bf C}^{\dagger}{\bf X}_j = \sum_{\ell}\left(2\ell +1 \right)\frac{\hat{\sigma}^{ij}_{\ell}}{C^{\prime}_{\ell}}\, ,
\end{eqnarray}  
where 
\begin{eqnarray}
\hat{\sigma}^{ij}_{\ell} = \sum_{m=-\ell}^{\ell} a^i_{\ell m} a^{j*}_{\ell m}/\left(2\ell+1\right)\, . 
\end{eqnarray}

\section{B: Conditional Density of CMB Theoretical Angular Power Spectrum}
\label{B}
We note that, if $x_1$, $x_2$, ..., $x_{\mu}$ are identically distributed and independent Gaussian random 
variable with  zero mean and unit variance, the new variable $x = \sum_{k=1}^{\mu}x^2_k$ is distributed 
as a $\chi^2$ random variable with $\mu$ degrees of freedom, with the probability density given by, 
\begin{eqnarray}
P(x) = \frac{1}{2^{\nu}\Gamma(\nu)} x^{\nu-1}\exp\left[ -\frac{x}{2}\right]\, , 
\label{chi2}
\end{eqnarray} 
where $\nu = \mu/2$. With this definition, the variable $x \equiv \left(2\ell +1\right) \frac{\hat C_{\ell}}{C_{\ell}}$
is distributed as Eqn.~\ref{chi2}, where $\hat C_{\ell}$  and $ C_{\ell}$ respectively denote realization specific and theoretical 
CMB angular power spectrum. To find the density function of $\hat C_{\ell}$ we first note using Eqn.~\ref{chi2} that, the density function $Q(y)$ 
for the transformed  variable $y = \beta x $ ($\beta$ = constant) follows, $Q(y) = P(x)dx/dy$, where in the right hand side $x$ must be 
replaced by $y$ using the inverse transformation $x = y/\beta$, so that one gets a function of $y$ as required. Using this concept 
and defining $y \equiv \hat C_{\ell} = C_{\ell}x/(2\ell+1)$, so that, $\beta = C_{\ell}/(2\ell +1)$, we find, 
\begin{eqnarray}
Q(\hat C_{\ell})= \left[2^{\left(2\ell+1\right)/2}\Gamma\left(\frac{2\ell+1}{2}\right)\right]^{-1}
\left(\frac{2\ell+1}{C_{\ell}}\right)^{\left(2\ell+1\right)/2} {\hat C}^{\left(2\ell+1\right)/2-1}_{\ell}
\exp\left[-\frac{\hat C_{\ell}\left(2\ell+1\right)}{2C_{\ell}}\right]
\label{pdf_clhat}
\end{eqnarray}   
Assuming $C_{\ell}$ as a random variable Eqn.~\ref{pdf_clhat} represents the conditional probability density
 $Q\left(\hat C_{\ell}|C_{\ell}\right)$. Using Bayes theorem and an uniform prior on $C_{\ell}$ upto some irrelevant 
constant, probability density of $C_{\ell}$ given some $\hat C_{\ell}$ can be obtained as, 
\begin{eqnarray}
R\left( C_{\ell}|\hat C_{\ell}\right) \propto \left(\frac{1}{C_{\ell}}\right)^{\left(2\ell+1\right)/2}
\exp\left[-\frac{\hat C_{\ell}\left(2\ell+1\right)}{2C_{\ell}}\right] 
\label{pdf_cl}
\end{eqnarray}
Now defining a new variable $z = {\hat C_{\ell}\left(2\ell+1\right)}/{C_{\ell}}$ and noting that the exponent of $1/C_{\ell}$ in   
Eqn.~\ref{pdf_cl} can be written as $(2\ell +1)/2 = (2\ell-1)/2 +1$ we can write Eqn.~\ref{pdf_cl} as, 
\begin{eqnarray}
R\left (z|\hat C_{\ell}\right) \propto z^{-(2\ell-1)/2 -1}\exp\left[-\frac{z}{2}\right]\, ,
\label{z_pdf} 
\end{eqnarray}     
where we have omitted some irrelevant constants. Comparing Eqn.~\ref{z_pdf} with Eqn.~\ref{chi2} we readily identify 
Eqn.~\ref{z_pdf} as a $\chi^2$ distribution of $2\ell-1$ degrees of freedom in variable $z$.  

\begin{thebibliography}{}
\expandafter\ifx\csname natexlab\endcsname\relax\def\natexlab#1{#1}\fi

\bibitem[{{Basak} \& {Delabrouille}(2012)}]{Delabrouille2012}
{Basak}, S., \& {Delabrouille}, J. 2012, \mnras, 419, 1163

\bibitem[{{Basak} \& {Delabrouille}(2013)}]{Delabrouille2013}
---. 2013, \mnras, 435, 18

\bibitem[{{Bennett} {et~al.}(1992){Bennett}, {Smoot}, {Hinshaw}, {Wright},
  {Kogut}, {de Amici}, {Meyer}, {Weiss}, {Wilkinson}, {Gulkis}, {Janssen},
  {Boggess}, {Cheng}, {Hauser}, {Kelsall}, {Mather}, {Moseley}, {Murdock}, \&
  {Silverberg}}]{Bennett1992}
{Bennett}, C.~L., {Smoot}, G.~F., {Hinshaw}, G., {et~al.} 1992, \apjl, 396, L7

\bibitem[{{Bennett} {et~al.}(2003){Bennett}, {Hill}, {Hinshaw}, {Nolta},
  {Odegard}, {Page}, {Spergel}, {Weiland}, {Wright}, {Halpern}, {Jarosik},
  {Kogut}, {Limon}, {Meyer}, {Tucker}, \& {Wollack}}]{Bennett2003}
{Bennett}, C.~L., {Hill}, R.~S., {Hinshaw}, G., {et~al.} 2003, \apjs, 148, 97

\bibitem[{{Bennett} {et~al.}(2013){Bennett}, {Larson}, {Weiland}, {Jarosik},
  {Hinshaw}, {Odegard}, {Smith}, {Hill}, {Gold}, {Halpern}, {Komatsu}, {Nolta},
  {Page}, {Spergel}, {Wollack}, {Dunkley}, {Kogut}, {Limon}, {Meyer}, {Tucker},
  \& {Wright}}]{Bennett2013}
{Bennett}, C.~L., {Larson}, D., {Weiland}, J.~L., {et~al.} 2013, \apjs, 208, 20

\bibitem[{{Bouchet} \& {Gispert}(1999)}]{Francois1999}
{Bouchet}, F.~R., \& {Gispert}, R. 1999, \na, 4, 443

\bibitem[{{Bouchet} {et~al.}(1999){Bouchet}, {Prunet}, \&
  {Sethi}}]{Bouchet1999}
{Bouchet}, F.~R., {Prunet}, S., \& {Sethi}, S.~K. 1999, \mnras, 302, 663

\bibitem[{Brooks \& Gelman(1998)}]{Brooks1998}
Brooks, S.~P., \& Gelman, A. 1998, Journal of Computational and Graphical
  Statistics, 7, 434

\bibitem[{{Bunn} {et~al.}(1994){Bunn}, {Fisher}, {Hoffman}, {Lahav}, {Silk}, \&
  {Zaroubi}}]{Bunn1994}
{Bunn}, E.~F., {Fisher}, K.~B., {Hoffman}, Y., {et~al.} 1994, \apjl, 432, L75

\bibitem[{{Eriksen} {et~al.}(2004{\natexlab{a}}){Eriksen}, {Banday},
  {G{\'o}rski}, \& {Lilje}}]{Eriksen2004}
{Eriksen}, H.~K., {Banday}, A.~J., {G{\'o}rski}, K.~M., \& {Lilje}, P.~B.
  2004{\natexlab{a}}, \apj, 612, 633

\bibitem[{{Eriksen} {et~al.}(2008{\natexlab{a}}){Eriksen}, {Dickinson},
  {Jewell}, {Banday}, {G{\'o}rski}, \& {Lawrence}}]{Eriksen2008}
{Eriksen}, H.~K., {Dickinson}, C., {Jewell}, J.~B., {et~al.}
  2008{\natexlab{a}}, \apjl, 672, L87

\bibitem[{{Eriksen} {et~al.}(2008{\natexlab{b}}){Eriksen}, {Jewell},
  {Dickinson}, {Banday}, {G{\'o}rski}, \& {Lawrence}}]{Eriksen2008a}
{Eriksen}, H.~K., {Jewell}, J.~B., {Dickinson}, C., {et~al.}
  2008{\natexlab{b}}, \apj, 676, 10

\bibitem[{{Eriksen} {et~al.}(2004{\natexlab{b}}){Eriksen}, {O'Dwyer}, {Jewell},
  {Wandelt}, {Larson}, {G{\'o}rski}, {Levin}, {Banday}, \&
  {Lilje}}]{Eriksen2004a}
{Eriksen}, H.~K., {O'Dwyer}, I.~J., {Jewell}, J.~B., {et~al.}
  2004{\natexlab{b}}, \apjs, 155, 227

\bibitem[{{Eriksen} {et~al.}(2006){Eriksen}, {Dickinson}, {Lawrence},
  {Baccigalupi}, {Banday}, {G{\'o}rski}, {Hansen}, {Lilje}, {Pierpaoli},
  {Seiffert}, {Smith}, \& {Vanderlinde}}]{Eriksen2006}
{Eriksen}, H.~K., {Dickinson}, C., {Lawrence}, C.~R., {et~al.} 2006, \apj, 641,
  665

\bibitem[{{Eriksen} {et~al.}(2007){Eriksen}, {Huey}, {Saha}, {Hansen}, {Dick},
  {Banday}, {G{\'o}rski}, {Jain}, {Jewell}, {Knox}, {Larson}, {O'Dwyer},
  {Souradeep}, \& {Wandelt}}]{Eriksen2007}
{Eriksen}, H.~K., {Huey}, G., {Saha}, R., {et~al.} 2007, \apj, 656, 641

\bibitem[{Gelman \& Rubin(1992)}]{GR1992}
Gelman, A., \& Rubin, D. 1992, Statistical Science, 1, 457

\bibitem[{Geman \& Geman(1984)}]{Gibbs1984}
Geman, S., \& Geman, D. 1984, IEEE Trans. Pattern Anal. Mach. Intell., 6, 721

\bibitem[{{Gold} {et~al.}(2011){Gold}, {Odegard}, {Weiland}, {Hill}, {Kogut},
  {Bennett}, {Hinshaw}, {Chen}, {Dunkley}, {Halpern}, {Jarosik}, {Komatsu},
  {Larson}, {Limon}, {Meyer}, {Nolta}, {Page}, {Smith}, {Spergel}, {Tucker},
  {Wollack}, \& {Wright}}]{Gold2011}
{Gold}, B., {Odegard}, N., {Weiland}, J.~L., {et~al.} 2011, \apjs, 192, 15

\bibitem[{{G{\'o}rski} {et~al.}(2005){G{\'o}rski}, {Hivon}, {Banday},
  {Wandelt}, {Hansen}, {Reinecke}, \& {Bartelmann}}]{Gorski2005}
{G{\'o}rski}, K.~M., {Hivon}, E., {Banday}, A.~J., {et~al.} 2005, \apj, 622,
  759

\bibitem[{{Hinshaw} {et~al.}(2013){Hinshaw}, {Larson}, {Komatsu}, {Spergel},
  {Bennett}, {Dunkley}, {Nolta}, {Halpern}, {Hill}, {Odegard}, {Page}, {Smith},
  {Weiland}, {Gold}, {Jarosik}, {Kogut}, {Limon}, {Meyer}, {Tucker}, {Wollack},
  \& {Wright}}]{Hinshaw2013}
{Hinshaw}, G., {Larson}, D., {Komatsu}, E., {et~al.} 2013, \apjs, 208, 19

\bibitem[{Moore(1920)}]{Moore1920}
Moore, E.~H. 1920, Bull. Am. Math. Soc., 26, 394, unpublished address.
  Available at
  http://www.ams.org/journals/bull/1920-26-09/S0002-9904-1920-03322-7/S0002-9904-1920-03322-7.pdf.

\bibitem[{Penrose(1955)}]{Penrose1955}
Penrose, R. 1955, Mathematical Proceedings of the Cambridge Philosophical
  Society, 51, 406

\bibitem[{{Penzias} \& {Wilson}(1965)}]{P&W1965}
{Penzias}, A.~A., \& {Wilson}, R.~W. 1965, \apj, 142, 419

\bibitem[{{Planck Collaboration} {et~al.}(2016{\natexlab{a}}){Planck
  Collaboration}, {Adam}, {Ade}, {Aghanim}, {Arnaud}, {Ashdown}, {Aumont},
  {Baccigalupi}, {Banday}, {Barreiro}, \& et~al.}]{PlanckCMB2016}
{Planck Collaboration}, {Adam}, R., {Ade}, P.~A.~R., {et~al.}
  2016{\natexlab{a}}, \aap, 594, A9

\bibitem[{{Planck Collaboration} {et~al.}(2016{\natexlab{b}}){Planck
  Collaboration}, {Adam}, {Ade}, {Aghanim}, {Arnaud}, {Ashdown}, {Aumont},
  {Baccigalupi}, {Banday}, {Barreiro}, \& et~al.}]{Planck2016_CMB}
---. 2016{\natexlab{b}}, \aap, 594, A9

\bibitem[{{Planck Collaboration} {et~al.}(2016{\natexlab{c}}){Planck
  Collaboration}, {Adam}, {Ade}, {Aghanim}, {Alves}, {Arnaud}, {Ashdown},
  {Aumont}, {Baccigalupi}, {Banday}, \& et~al.}]{PlanckFg2016}
---. 2016{\natexlab{c}}, \aap, 594, A10

\bibitem[{{Planck Collaboration} {et~al.}(2016{\natexlab{d}}){Planck
  Collaboration}, {Ade}, {Aghanim}, {Arnaud}, {Ashdown}, {Aumont},
  {Baccigalupi}, {Banday}, {Barreiro}, {Bartlett}, \&
  et~al.}]{PlanckCosmoParam2016}
{Planck Collaboration}, {Ade}, P.~A.~R., {Aghanim}, N., {et~al.}
  2016{\natexlab{d}}, \aap, 594, A13

\bibitem[{{Planck Collaboration} {et~al.}(2018{\natexlab{a}}){Planck
  Collaboration}, {Akrami}, {Arroja}, {Ashdown}, {Aumont}, {Baccigalupi},
  {Ballardini}, {Banday}, {Barreiro}, {Bartolo}, {Basak}, {Battye}, {Benabed},
  {Bernard}, {Bersanelli}, {Bielewicz}, {Bock}, {Bond}, {Borrill}, {Bouchet},
  {Boulanger}, {Bucher}, {Burigana}, {Butler}, {Calabrese}, {Cardoso},
  {Carron}, {Casaponsa}, {Challinor}, {Chiang}, {Colombo}, {Combet},
  {Contreras}, {Crill}, {Cuttaia}, {de Bernardis}, {de Zotti}, {Delabrouille},
  {Delouis}, {D{\'e}sert}, {Di Valentino}, {Dickinson}, {Diego}, {Donzelli},
  {Dor{\'e}}, {Douspis}, {Ducout}, {Dupac}, {Efstathiou}, {Elsner},
  {En{\ss}lin}, {Eriksen}, {Falgarone}, {Fantaye}, {Fergusson},
  {Fernandez-Cobos}, {Finelli}, {Forastieri}, {Frailis}, {Franceschi},
  {Frolov}, {Galeotta}, {Galli}, {Ganga}, {G{\'e}nova-Santos}, {Gerbino},
  {Ghosh}, {Gonz{\'a}lez-Nuevo}, {G{\'o}rski}, {Gratton}, {Gruppuso},
  {Gudmundsson}, {Hamann}, {Handley}, {Hansen}, {Helou}, {Herranz}, {Hivon},
  {Huang}, {Jaffe}, {Jones}, {Karakci}, {Keih{\"a}nen}, {Keskitalo}, {Kiiveri},
  {Kim}, {Kisner}, {Knox}, {Krachmalnicoff}, {Kunz}, {Kurki-Suonio}, {Lagache},
  {Lamarre}, {Langer}, {Lasenby}, {Lattanzi}, {Lawrence}, {Le Jeune}, {Leahy},
  {Lesgourgues}, {Levrier}, {Lewis}, {Liguori}, {Lilje}, {Lilley}, {Lindholm},
  {L{\'o}pez-Caniego}, {Lubin}, {Ma}, {Mac{\'{\i}}as-P{\'e}rez}, {Maggio},
  {Maino}, {Mandolesi}, {Mangilli}, {Marcos-Caballero}, {Maris}, {Martin},
  {Mart{\'{\i}}nez-Gonz{\'a}lez}, {Matarrese}, {Mauri}, {McEwen}, {Meerburg},
  {Meinhold}, {Melchiorri}, {Mennella}, {Migliaccio}, {Millea}, {Mitra},
  {Miville-Desch{\^e}nes}, {Molinari}, {Moneti}, {Montier}, {Morgante}, {Moss},
  {Mottet}, {M{\"u}nchmeyer}, {Natoli}, {N{\o}rgaard-Nielsen}, {Oxborrow},
  {Pagano}, {Paoletti}, {Partridge}, {Patanchon}, {Pearson}, {Peel}, {Peiris},
  {Perrotta}, {Pettorino}, {Piacentini}, {Polastri}, {Polenta}, {Puget},
  {Rachen}, {Reinecke}, {Remazeilles}, {Renzi}, {Rocha}, {Rosset}, {Roudier},
  {Rubi{\~n}o-Mart{\'{\i}}n}, {Ruiz-Granados}, {Salvati}, {Sandri},
  {Savelainen}, {Scott}, {Shellard}, {Shiraishi}, {Sirignano}, {Sirri},
  {Spencer}, {Sunyaev}, {Suur-Uski}, {Tauber}, {Tavagnacco}, {Tenti},
  {Terenzi}, {Toffolatti}, {Tomasi}, {Trombetti}, {Valiviita}, {Van Tent},
  {Vibert}, {Vielva}, {Villa}, {Vittorio}, {Wandelt}, {Wehus}, {White},
  {White}, {Zacchei}, \& {Zonca}}]{Planckcosmo2018}
{Planck Collaboration}, {Akrami}, Y., {Arroja}, F., {et~al.}
  2018{\natexlab{a}}, ArXiv e-prints, arXiv:1807.06205

\bibitem[{{Planck Collaboration} {et~al.}(2018{\natexlab{b}}){Planck
  Collaboration}, {Akrami}, {Ashdown}, {Aumont}, {Baccigalupi}, {Ballardini},
  {Banday}, {Barreiro}, {Bartolo}, {Basak}, {Benabed}, {Bersanelli},
  {Bielewicz}, {Bond}, {Borrill}, {Bouchet}, {Boulanger}, {Bucher}, {Burigana},
  {Calabrese}, {Cardoso}, {Carron}, {Casaponsa}, {Challinor}, {Colombo},
  {Combet}, {Crill}, {Cuttaia}, {de Bernardis}, {de Rosa}, {de Zotti},
  {Delabrouille}, {Delouis}, {Di Valentino}, {Dickinson}, {Diego}, {Donzelli},
  {Dor{\'e}}, {Ducout}, {Dupac}, {Efstathiou}, {Elsner}, {En{\ss}lin},
  {Eriksen}, {Falgarone}, {Fernandez-Cobos}, {Finelli}, {Forastieri},
  {Frailis}, {Fraisse}, {Franceschi}, {Frolov}, {Galeotta}, {Galli}, {Ganga},
  {G{\'e}nova-Santos}, {Gerbino}, {Ghosh}, {Gonz{\'a}lez-Nuevo}, {G{\'o}rski},
  {Gratton}, {Gruppuso}, {Gudmundsson}, {Handley}, {Hansen}, {Helou},
  {Herranz}, {Huang}, {Jaffe}, {Karakci}, {Keih{\"a}nen}, {Keskitalo},
  {Kiiveri}, {Kim}, {Kisner}, {Krachmalnicoff}, {Kunz}, {Kurki-Suonio},
  {Lagache}, {Lamarre}, {Lasenby}, {Lattanzi}, {Lawrence}, {Le Jeune},
  {Levrier}, {Liguori}, {Lilje}, {Lindholm}, {L{\'o}pez-Caniego}, {Lubin},
  {Ma}, {Mac{\'{\i}}as-P{\'e}rez}, {Maggio}, {Maino}, {Mandolesi}, {Mangilli},
  {Marcos-Caballero}, {Martin}, {Mart{\'{\i}}nez-Gonz{\'a}lez}, {Matarrese},
  {Mauri}, {McEwen}, {Meinhold}, {Melchiorri}, {Mennella}, {Migliaccio},
  {Miville-Desch{\^e}nes}, {Molinari}, {Moneti}, {Montier}, {Morgante},
  {Natoli}, {Oppizzi}, {Pagano}, {Paoletti}, {Partridge}, {Peel}, {Pettorino},
  {Piacentini}, {Polenta}, {Puget}, {Rachen}, {Reinecke}, {Remazeilles},
  {Renzi}, {Rocha}, {Roudier}, {Rubi{\~n}o-Mart{\'{\i}}n}, {Ruiz-Granados},
  {Salvati}, {Sandri}, {Savelainen}, {Scott}, {Seljebotn}, {Sirignano},
  {Spencer}, {Suur-Uski}, {Tauber}, {Tavagnacco}, {Tenti}, {Thommesen},
  {Toffolatti}, {Tomasi}, {Trombetti}, {Valiviita}, {Van Tent}, {Vielva},
  {Villa}, {Vittorio}, {Wandelt}, {Wehus}, {Zacchei}, \& {Zonca}}]{Planck2018}
{Planck Collaboration}, {Akrami}, Y., {Ashdown}, M., {et~al.}
  2018{\natexlab{b}}, ArXiv e-prints, arXiv:1807.06208

\bibitem[{{Purkayastha} \& {Saha}(2017)}]{Ujjal2017}
{Purkayastha}, U., \& {Saha}, R. 2017, ArXiv e-prints, arXiv:1707.02008

\bibitem[{{Saha}(2011)}]{Saha2011}
{Saha}, R. 2011, \apjl, 739, L56

\bibitem[{{Saha} \& {Aluri}(2016)}]{Saha2016}
{Saha}, R., \& {Aluri}, P.~K. 2016, \apj, 829, 113

\bibitem[{{Saha} {et~al.}(2006){Saha}, {Jain}, \& {Souradeep}}]{Saha2006}
{Saha}, R., {Jain}, P., \& {Souradeep}, T. 2006, \apjl, 645, L89

\bibitem[{{Saha} {et~al.}(2008){Saha}, {Prunet}, {Jain}, \&
  {Souradeep}}]{Saha2008}
{Saha}, R., {Prunet}, S., {Jain}, P., \& {Souradeep}, T. 2008, \prd, 78, 023003

\bibitem[{{Smoot} {et~al.}(1991){Smoot}, {Bennett}, {Kogut}, {Aymon}, {Backus},
  {de Amici}, {Galuk}, {Jackson}, {Keegstra}, {Rokke}, {Tenorio}, {Torres},
  {Gulkis}, {Hauser}, {Janssen}, {Mather}, {Weiss}, {Wilkinson}, {Wright},
  {Boggess}, {Cheng}, {Kelsall}, {Lubin}, {Meyer}, {Moseley}, {Murdock},
  {Shafer}, \& {Silverberg}}]{COBE1991}
{Smoot}, G.~F., {Bennett}, C.~L., {Kogut}, A., {et~al.} 1991, \apjl, 371, L1

\bibitem[{{Sudevan} {et~al.}(2017){Sudevan}, {Aluri}, {Yadav}, {Saha}, \&
  {Souradeep}}]{Sudevan2017}
{Sudevan}, V., {Aluri}, P.~K., {Yadav}, S.~K., {Saha}, R., \& {Souradeep}, T.
  2017, \apj, 842, 62

\bibitem[{{Sudevan} \& {Saha}(2017)}]{Saha2017}
{Sudevan}, V., \& {Saha}, R. 2017, ArXiv e-prints, arXiv:1712.09804

\bibitem[{{Tegmark} {et~al.}(2003){Tegmark}, {de Oliveira-Costa}, \&
  {Hamilton}}]{Tegmark2003}
{Tegmark}, M., {de Oliveira-Costa}, A., \& {Hamilton}, A.~J. 2003, Phys. Rev.
  D, 68, 123523

\bibitem[{{Tegmark} \& {Efstathiou}(1996)}]{Tegmark96}
{Tegmark}, M., \& {Efstathiou}, G. 1996, Mon. Not. R. Astron. Soc., 281, 1297

\end{thebibliography}

\end{document}